# Tilt-to-length coupling metrology in the LISA mission

Frédéric Cleva[1a], Jean-Pierre Coulon [a], Marco Nardello [a]

[a]Université Côte d'Azur, Observatoire de la Côte d'Azur, CNRS, ARTEMIS, NICE, France

**ABSTRACT**

This paper describes a setup aimed at measuring the so-called Tilt-To-Length (TTL) coupling in the optical benches of the LISA mission. The TTL is the coupling of the angular jitter of any optical setup into the optical path length between its input and output pupils. This might be deleterious in laser ranging experiments and must be evaluated for further compensation. The setup is made of two laser beams, one features an angular jitter that mimics the input beam as seen from the jittering bench under test (BUT), the other is aligned to the optical axis of the BUT and provides a phase reference for the jittering beam. The induced phase variations between both beams detected at the BUT's output pupil gives access to the TTL coupling. The "TTL probe" must feature a negligible residual TTL coupling which implies a micrometric accuracy in the centering of the setup pupil, the beams and the angular jitter associated pivot point. The setup integrates optical masks as a link between the setup optical reference frame to its mechanical reference frame, together with position memories and servo-loops for the beam's alignment. We show that the stability, the accuracy, and the noise floor of the setup is compliant with the LISA specifications for the TTL mitigation, although it makes use of off-the-shelf components and is operated in a standard environment laboratory.

## 1. INTRODUCTION

The first detection of a Gravitational Wave (GW) by terrestrial gravitational waves detector [1] has opened the path for a new GW astronomy. On early 2024 more than 171 events have been routinely detected by the Virgo and LIGO terrestrial detectors [2] [3]. Within this panorama the Laser Interferometer Space Antenna mission, LISA [4], is the project for a spatial interferometric gravitational waves detector which aims at the detection in a lower frequency range, inaccessible to the terrestrial detectors because of earth related showstoppers[2] [3]. LISA is an ESA project with NASA collaboration. It has been adopted in January 2024, the launch is foreseen by 2037.

LISA will detect GWs by measuring the distance variations between free falling masses (FFM), which are induced by the space-time structure distortions the GWs are made of. LISA will be able to detect Optical Path Length (OPL) variations of less than $10\ pm/\sqrt{Hz}$ in the range [0.1 mHz – 1 Hz]. The FFMs are separated by 2.5 Mkm distance and connected through an optical laser link (1.064 um wavelength). The variation of the accumulated optical phase all along the link contains the GW signature and the detection is made through the interferometric combination of the incoming laser link (so-called "Rx" beam) and a laser attached to the local FFM (so-called "LO" beam). The FFMs are surrounded by an optical bench (LISA_OB) which holds the photodetectors and integrates the optical setup which manages the beams combination. One of the key issues for LISA comes from the coupling of the angular jittering of the bench, triggered by the satellite residual movements, with longitudinal OPL in case of the output-to-input bench pupils misalignments. The extra OPL will end up spoiling the science signal. This deleterious effect is the Tilt-To-Length coupling (TTL) and must be precisely measured to be able to compensate for it. We have developed a setup aiming at the measurement of the TTL coupling compliant with what is required for LISA optical benches. This paper describes some key features and the performances of this demonstrator, the so-called FOGOB.

---

[1] Frederic.cleva@oca.eu, phone: +33 (0)4 92 00 31 97
[2]  0.1 mHz to 1 Hz for LISA, and 10 Hz to 10 kHz for Virgo/LIGO
[3] Seismic and thermal noise, Newtonian noise …

## 2. FOGOB CONCEPTS

### 2.1. Operating principle

The TTL coupling can be understood as a leverage effect, it is expressed in m/radian as $TTL = \frac{\partial \delta x}{\partial \theta} = \Delta$, where $\delta x$ is the longitudinal OPL change induced by the tilt of the optical bench under tests and "$\Delta$" is the offset of the incident Rx beam axis with respect to the tilt pivot point (Figure 1). The FOGOB delivers two laser beams which mimic the Rx and the LO beams. The LO beam is aligned with the optical axis of the LISA_OB, while the Rx beam mimics the far incoming optical link[4]. The Rx and LO beams axis are superimposed on each other and phase-locked at the center of the FOGOB output pupil, which insures a null TTL coupling at this location. The FOGOB output pupil is conjugated with the pupil of the LISA_OB (see Figure 2). In this configuration the TTL coupling at the LISA_OB is deduced from the relative phase variations accumulated all along the propagation of the Rx beam through the LISA_OB with respect to the LO beam. The measurement is made at the LISA_OB photodetector.

### 2.2. Specifications

The FOGOB must be able to sense the TTL coupling coefficient with an accuracy better than 15 µm/radian and with a precision better than 1 µm/radian at 3 σ (statistics assessed during the measurement campaigns, ~ 30 minutes) and all along the operational period (few days) before any calibration of the setup.

### 2.3. Description

The Rx beam features a sinusoidal angular jitter at a frequency of a few tens of Hz, with the associated "pivot point" centered on the FOGOB pupil (Figure 2). All the centering's are achieved with a micrometric accuracy and kept for long periods (days). We use a heterodyne detection scheme with a 1 MHz frequency shift between the Rx and LO beams which ends up as a 1 MHz beat at the photodetector. After demodulation at 1 MHz, we get the phase related to the OPL according to the relation $\varphi = \frac{2.\pi}{\lambda} \times OPL$, ($\lambda$ = 1064 nm is the laser wavelength). The phase is measured over the whole pupil (2.55 mm FOGOB diameter aperture) with a 4-quadrants photodetector. The TTL is the sum of the phase variations detected at each quadrant, expressed in meters of OPL and normalized with the associated "Rx" tilt (Figure 3). The phase difference between pairs of adjacent quadrants gives the angle between the Rx and the LO beams, that is the so-called the Differential Wavefront Signals (DWS), for both the horizontal and the vertical directions (Figure 4).

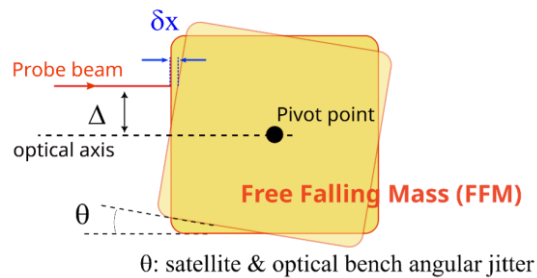

*Figure 1 illustrates the TTL coupling as a lever effect. The schematic refers to the satellite referential frame which detects the FFM as jittering.*

---

[4] In the reference frame of the LISA_OB the Rx beam is seen as angularly jittering

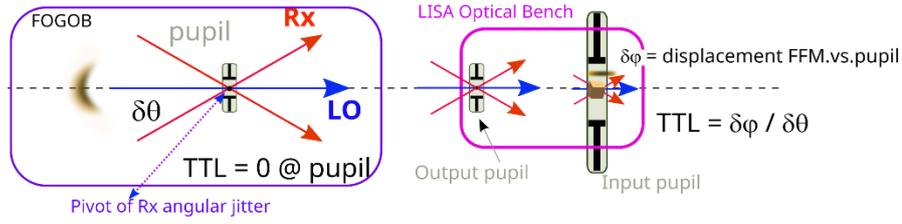

*Figure 2 illustrates the FOGOB principle. The FOGOB pupil is conjugated with the input and output pupils of the LISA_OB. The jittering FFM is set at the LISA_OB input pupil. The TTL coupling occurs due the misalignments among the LISA_OB input and output pupils and the FFM. In this sketch we have not considered the optical elements which conjugate the FOGOB pupil on the LISA_OB pupil and which are not part of this study.*

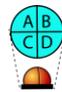

*Figure 3 TTL coupling coefficient as measured with 4 Quadrants Photodetector (QPD)*

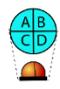

*Figure 4: the Differential Wavefront Signal (DWS) signal definition. DWS accounts for the tilt between Rx and LO beams along the vertical and the horizontal directions. The right scheme illustrates the vertical related DWS.*

## 3. FOGOB INTEGRATION

### 3.1. OVERALL STRATEGY

As shown in the Figure 2 the FOGOB setup requires 3 key features:

- The Rx and LO beams axes are centered on the FOGOB pupil
- The Rx angular jitter pivot point is centered on the FOGOB pupil
- The Rx-to-LO is phase-locked at the FOGOB pupil center

We have used a strategy that has been setup at the Albert Einstein Institute in Hanover [5] [6]. It integrates 3 transmissive optical masks each sequentially placed at the same location during the three-steps FOGOB integration. Each mask is designed with a pattern devoted to either aligning the beams' axes at the center of the mask, defining the pupil (size[5] and positioning at the center of the mask), and defining the phase lock area (size[6] and positioning at the center of the mask). Figure 5 displays the three masks. The masks also integrate four "fiducial" which make the link between the pattern's reference frame and the FOGOB's mechanical frame. The fiducials enable an identical positioning of the three masks, which ends up with the centering the pivot point, the beams' axes, the pupil pattern on a common center which defines at the end the FOGOB pupil.

---

[5] 2.55 mm diameter aperture
[6] 150 microns diameter aperture

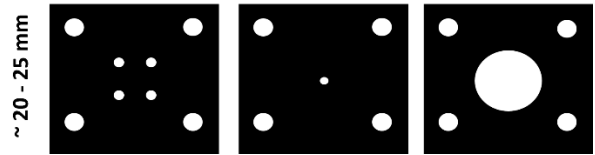

*Figure 5 Masks zoology: « QPD-like » mask for the beam centering, « Pinhole » mask for the phase-lock area definition (150 microns diameter), "Pupil" mask for the FOGOB pupil definition (2.55 mm diameter). The four holes at the mask's corners are the fiducials used to relay the pattern reference frame to the FOGOB reference frame*

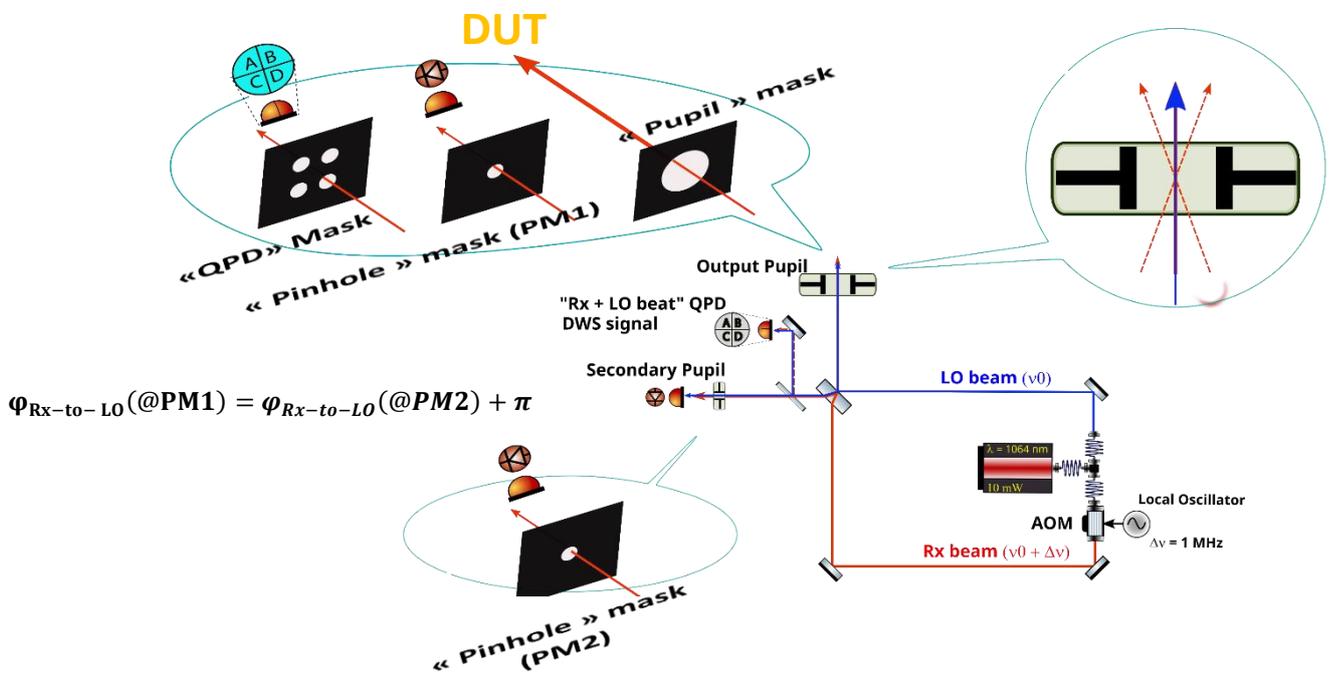

*Figure 6 : Masks-based strategy for the FOGOB integration. The "Rx+LO beat QPD" delivers the DWS signal after the beat demodulation with the 1 MHz local oscillator. Such operation has not been sketched in order to ease the reading.*

The overall strategy is sketched in the Figure 6.

- We align the beams on the "QPD" mask pattern by balancing the power transmitted at each hole of the 4-holes pattern. This sets the pattern center of the FOGOB optical frame.
- We replace the "QPD" mask with the "Pinhole" mask ("PM1"). We phase-lock the Rx to the LO beams with the error signal extracted from the 1 MHz beat detected at the pinhole transmission. We detect the error signal of a twin "pinhole mask + photodetector" assembly (PM2, at the secondary pupil, Figure 6) which is conjugated with PM1 with respect to the beamsplitter. Such a signal equals the former one (@PM1) modulo π. Last, we feed the phase-lock loop with the "twin" error signal.
- We replace the "Pinhole" mask (PM1) with the pupil mask, making the path free for Rx and LO toward the LISA_OB.

We end up with a configuration where the beams, the pivot point, the pupil are all centered and the Rx and LO beams phase-locked each other [5].

In the following we describe the setup we have developed for the management of the masks' positioning, the beam alignment control, and we show the overall performances of the FOGOB in terms of its residual TTL coupling.

### 3.2. MASKS' ALIGNEMENT

The masks are quartz substrates (28 x 28 x 1 mm), they are anti-reflective coated on both sides and black coated[7] on the whole surface but the areas which define the patterns of interest (Figure 7). The black coating is deposited through photolithography which insures the pattern design within a precision better than 0.25 µm. We have measured the deviations in sizes and distances of the various figures within the overall pattern to be smaller than 0.5 µm (compliant with the Coordinate Measurement Machine precision we have used).

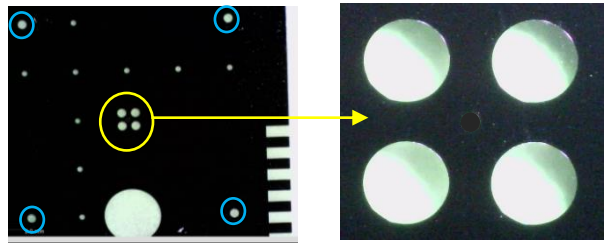

*Figure 7 shows an example of the mask devoted to the beam alignment. The patterns are defined by the uncoated areas. The yellow circle pinpoints the so-called "QPD-like" pattern which defines a reference frame for the beams centering, see §3.3. The blue circles pinpoint four fiducials (500-microns diameter disks) which link the reference frame of the "QPD-like" pattern to the FOGOB mechanical reference frame. The right picture is a zoom of the central pattern. The other patterns serve for metrology purposes. See also Figure 5*

We have designed an opto-mechanical setup which detects the mask's fiducials position within the FOGOB mechanical reference frame (Figure 8). It serves as a position memory which is required during the mask's replacement process. The positions of the 4 fiducials are detected by a set of 4 QPDs fixed in front of the masks, on an invar[8] support attached to the FOGOB bench. Each of the fiducial aperture is illuminated with a laser beam attached to the FOGOB (four fibered collimators, Figure 8) whose diffraction pattern is detected by the QPD, giving the fiducial aperture position. With such a configuration we were able to detect fiducial/mask displacements as small as 30 nm. The mask's support can slide on four thin sapphire pads glued on the support attached to the FOGOB. It is bounded to the FOGOB support thanks to 4 embedded magnets[9]. Such an assembly allows the disengagement of the actuators and avoids that any mechanical perturbations of the actuators couple in the mask support position. The mask position is finely adjusted with four micrometric horizontal stages (Figure 8)Figure 8. A vertical lever can raise the mask support if needed. Although such an assembly turned out to be quite tedious to use it has been chosen for cost reasons. More efficient and expensive solutions exist but are out of the scope of our demonstrator.

---

[7] Optical density > 4
[8] Invar is a FeNi36 alloy which shows a low thermal expansion coefficient
[9] The mask support is made of invar which shows magnetic properties

The stability of the mask position in the FOGOB reference frame is displayed in Figure 9. Two fiducials are presented along the horizontal direction. They show a long-term drift of their position smaller than 0.4 μm over 120 hours and less than 0.1 microns for short term periods (hours). Such a setup is therefore compliant for the masks positioning within an accuracy better than 0.1 micron and the long-term monitoring of the position within 0.4 micron.

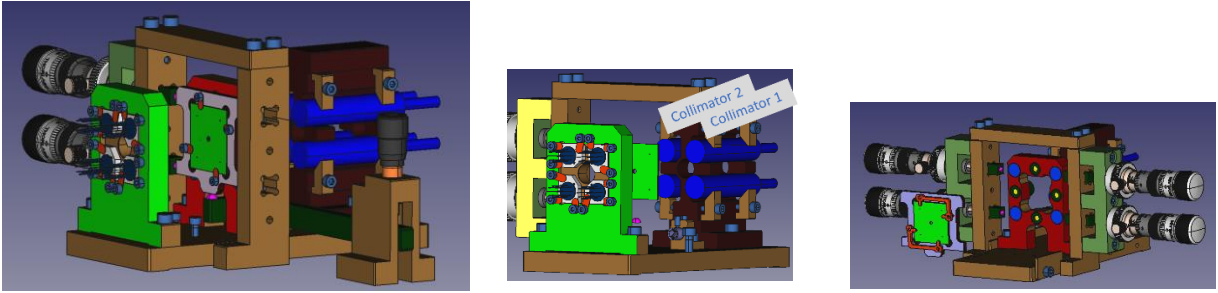

*Figure 8 "Position memory" setup for positioning the masks in the FOGOB mechanical reference frame. The masks support (magnetic invar) is bound to the red support attached to the FOGOB (left sketch), with four magnets (yellow disk, right scheme) and can slide (sledge-like) over 4 sapphire thin pads (blue disks) glued on the red support. The 4 fiducials "apertures" are visible near the corners of the mask. The Rx and LO beams propagate all along the setup thanks to a central aperture (middle sketch). Four horizontal micrometric stages are used to push-position the mask "sledge" (only two stages are shown for an easy reading of the design).*

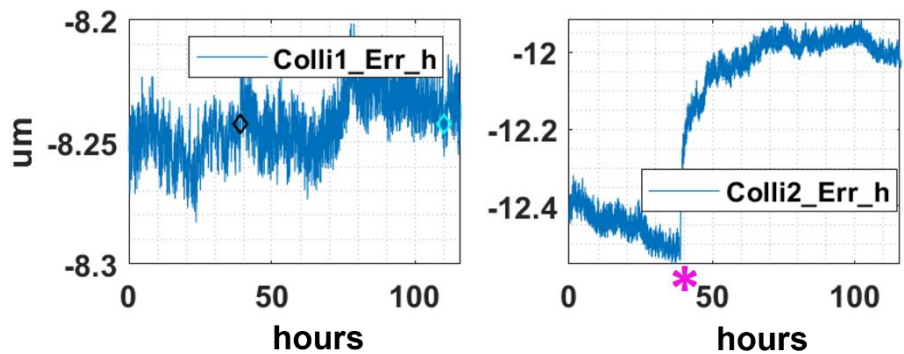

*Figure 9: The long-term monitor of the fiducials position associated with the collimators 1 and 2 (upper collimators). Only the horizontal drifts are shown. The pink star indicates an on-purpose perturbation of the setup; therefore, the sharp 0.4-micron drift should not be accounted for in the free running stability. The positions drift by less than 0.5 μm over 120 hours and less than 0.1 microns over few hours.*

### 3.3. LO BEAM ALIGNEMENT

We have implemented a strategy based on a "Position memory QPD" to finely tune the beam position at the FOGOB pupil. The use of a servo-loop which clamps the beam at the position memory QPD makes quite easy submicrometric tuning and improves the stability of the beam position.

We integrate a QPD which is conjugated to the FOGOB pupil with respect to the pickoff beamsplitter which splits the LO beam. A servo-loop clamps the LO position at the position memory QPD through a piezo driven mirror (Figure 10). Any input added to the servo-loop error signal is considered by it as an actual beam displacement, which, in turn, will be compensated for by the loop with an actual beam displacement. Such a method allows for quite fine and straightforward alignments within a submicrometric accuracy as long as we stay in the error signal linear regime. We use this actuation strategy to adjust the LO beam at the FOGOB pupil during the "QPD" mask integration (see 3). Once this is done the beam is kept aligned thanks to the servo and the bias introduced in the error signal.

Figure 11 (left plot), illustrates the effect of an on-purpose "perturbation" added to the servo-loop error signal which drives the LO beam position at the position memory QPD, with the same effect detected at the FOGOB output pupil (QPD-like mask). The light blue curve (position at the FOGOB pupil) features some extra noise likely due to some air "turbulence" in the laboratory.

The same strategy has been implemented for the control of the Rx beam. In this specific case we control four degrees of freedom: the beam tilt and shift along both horizontal and vertical directions. The error signals is given by a position memory QPD as for the LO beam configuration and by the DWS signal (§2.3, Figure 4) which monitors the angle between the Rx and the LO beams. We have diagonalized the 4 X 4 command matrix to disentangle the actuation on the 4 degrees of freedom of the system. We have reached the same performance as for the LO beam in terms of accuracy. Moreover, we were able to imprint a pure sinusoidal angular jitter to the Rx beam by simply add an appropriate sinusoidal perturbation to the DWS error signal. The Figure 11 (right plot) illustrates the effect a sinusoidal perturbation added at the DWS error signal input which ends up with a sinusoidal tilt visible at the DWS QPD and at a dedicated "tilt monitor QPD" [10].

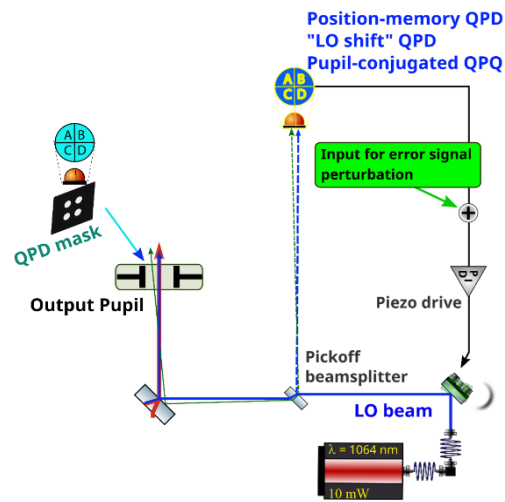

*Figure 10 Alignment servo-loop for the LO beam. The "LO shift" QPD is conjugated with the "QPD mask" at the output pupil. Once locked, the "QPD mask" can be removed while the "LO shift" QPD acts as a position memory for the LO positioning.*

---

[10] We use a QPD located at the focus of a lens, which is sensitive only to the tilt variations of the incoming beam.

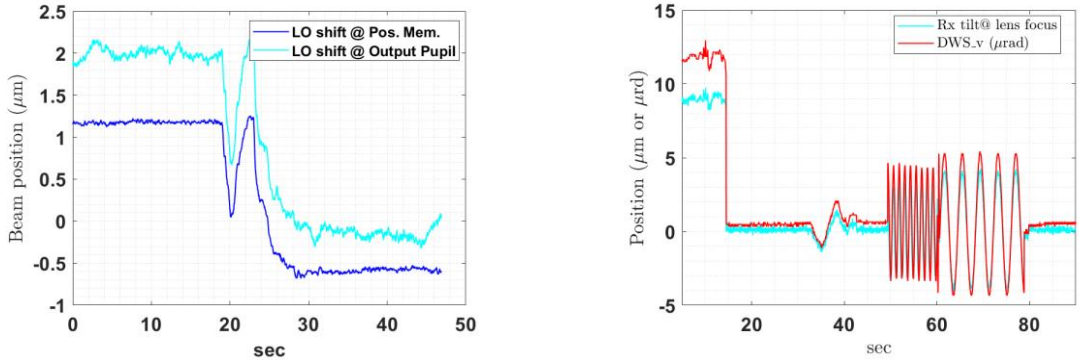

*Figure 11 : the left plot illustrates the effect of a specific signal added to the LO beam position error signal. The effect is visible at the "LO+Rx" QPD (Figure 10) which delivers the error signal and at the FOGOB pupil. The right plot illustrates the same effect for the Rx beam angular control. The loop is closed at t = 10 sec, a sharp perturbation is added at t = 35 sec., two sinusoidal perturbations are added after t = 50 sec, with 2 different frequencies.*

## 4. FOGOB RESIDUAL TTL COUPLING

We have measured the optical path length variations induced by an on-purpose sinusoidal angular jitter of the Rx beam (15 Hz). In a preliminary phase we have made the measurement at the "Rx + LO beat" QPD (Figure 6) rather than at the FOGOB pupil as it should be (Figure 2). In such a configuration the pupil is defined by the QPD aperture itself (3 mm diameter) instead of the FOGOB pupil (2.55 m diameter). Moreover, the measurement, so-called $TTL_{Meas.}$, differs from the TTL in that it deals with two pairs of merged quadrants instead of dealing with four single quadrants (Figure 12), so is for the DWS signals.

$$TTL_{Meas.} = \frac{2.\pi}{\lambda} \times \frac{1}{2} \times \frac{1}{\theta_0}(\varphi_{A+B} + \varphi_{C+D})$$

$$DWS_{vertical} \cong \varphi_{A+B} - \varphi_{C+D}$$

*Figure 12 Simplified expression for the TTL and DWS signals*

The phase signal is extracted from the 1 MHz beat after demodulation at the local oscillator frequency. We use a spectrum analyzer to monitor the amplitude of phase signal line at 15 Hz and express it in meter and get the TTL once normalized by the Rx angular jitter amplitude. The result is displayed in Figure 13.

The $TTL_{Meas.}$ shows a long-term drift smaller than 0.7 micron/radian (130 hours) and is assessed with an accuracy better than 5 micron/radian. It is therefore compliant with the specifications of 15 microns/radian accuracy and 1-micron/radian precision at 3 σ [11] (§2.2).

---

[11] The statistics over trends of 60 sec. data shows a standard deviation less than 0.1 micron/radian and decrease for longer trend.

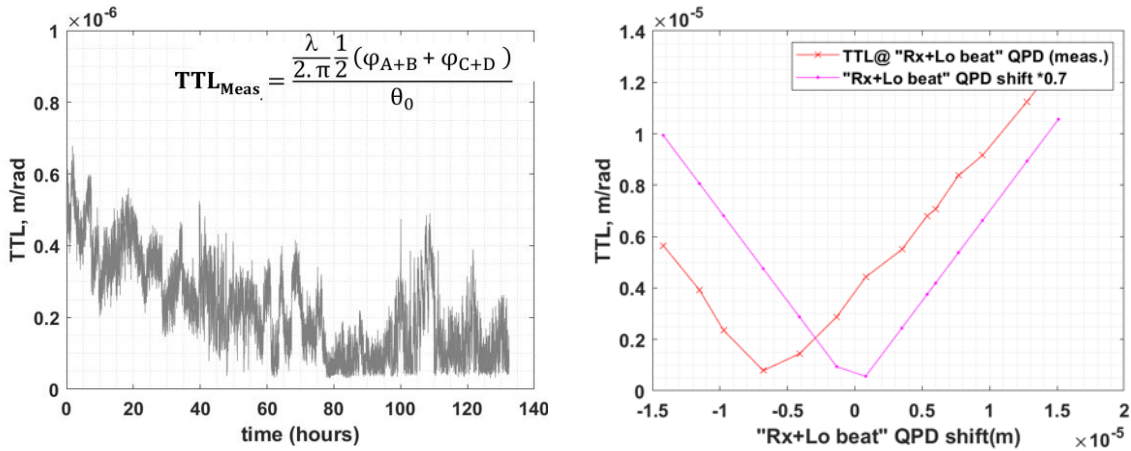

*Figure 13 The TTL$_{Meas.}$ trend over 130 hours (left plot). The TTL$_{Meas.}$ as a function of the QPD misalignment (right plot), the pink curve is given by a simulation. The residual TTL when the QPD is centered is ~ 4 μm/radian.*

## 5. CONCLUSION

We have developed the FOGOB setup aimed at the measurement of the TTL of devices like the optical bench of LISA. The accuracy of the TTL determination is better than 5 micron/radian, the precision better than 1 micron/radian over 130 hours. It makes use of off-shelf-components and is operated under normal environment conditions. The use of servo-loops and position memories to control the alignment of the beams makes it quite easy to achieve submicrometric alignments in a straightforward way and within short time. It may also be remotely controlled if needed. The servo-loops control copes with the possible issues related to the collimators non-qualified stability (at least for standard components). Such an unknown would erode the confidence we have on the stability of the setup, and hence on the accuracy of the residual TTL. The use of photolithography designed patterns and optical masks allows for a centering accuracy better than 1 micron. Although the definition of the measured TTL is a simplified version of the TTL definition LISA is concerned with, we are confident that the concepts developed for the FOGOB are compliant for the TTL measurements.

## ACKNOWLEDGEMENT

This work has been funded by the Centre National d'Etudes Spatiales (CNES).

## REFERENCES

[1] Abbott et al., "Observation of Gravitational Waves from a Binary Black Hole Merger," *Phys. Rev. Letters,* vol. 116, p. 061102, Feb. 2016.

[2] Virgo, «htttps://www.virgo-gw.eu/».

[3] LIGO, «https://www.ligo.caltech.edu/».

[4] «LISA - The Gravitational Universe,» [En ligne]. Available: https://sci.esa.int/web/lisa/-/58543-the-gravitational-universe.

[5] M. Chwalla, «Optical Suppression of Tilt-to-Length Coupling in the LISA Long-Arm Interferometer,» *Phys. Rev. Applied,* vol. 14, n° %11, p. 014030 , 2020.

[6] M. Tröbs, «Reducing tilt-to-length coupling for the LISA test mass interferometer,» *Class. Quantum Grav.,* vol. 35, n° %110, p. 105001, 2018.